\documentclass[RNAAS]{aastex63}

\usepackage{graphicx}
\usepackage{upgreek}
\graphicspath{{./}{figures/}}
\begin{document}

\title{Spectro-temporal analysis of a sample of bursts from FRB~121102}

\correspondingauthor{Kaustubh Rajwade}
\email{kaustubh.rajwade@manchester.ac.uk}

\author[0000-0002-8043-6909]{Kaustubh Rajwade}
\affiliation{University of Manchester, Oxford Road, Manchester M13 9PL, UK}

\author{Mitchell Mickaliger}
\affiliation{University of Manchester, Oxford Road, Manchester M13 9PL, UK}

\author[0000-0001-9242-7041]{Benjamin Stappers}
\affiliation{University of Manchester, Oxford Road, Manchester M13 9PL, UK}

\author[0000-0002-4079-4648]{Manisha Caleb}
\affiliation{University of Manchester, Oxford Road, Manchester M13 9PL, UK}


\author[0000-0001-8522-4983]{Rene Breton}
\affiliation{University of Manchester, Oxford Road, Manchester M13 9PL, UK}

\author[0000-0002-1434-9786]{Aris Karastergiou}
\affiliation{Astrophysics, Denys Wilkinson building, University of Oxford, Keble Road, Oxford OX1 3RH, UK}
\affiliation{Department of Physics and Electronics, Rhodes University, PO Box 94, Grahamstown 6140, South Africa}

\author[0000-0002-4553-655X]{Evan Keane}
\affiliation{SKA Organisation, Jodrell Bank, Macclesfield SK11 9FT, UK}
\affiliation{University of Manchester, Oxford Road, Manchester M13 9PL, UK}


\begin{abstract}

FRB~121102 was the first Fast Radio Burst (FRB) that was shown to repeat. Since its discovery in 2012, more than two hundred bursts have been detected from the source. These bursts exhibit a diverse range of spectral and temporal characteristics and many questions about their origin and form remain unanswered. Here, we present a sample of radio bursts from FRB 121102 detected using the Lovell telescope at Jodrell Bank Observatory. We show four examples of bursts that show peculiar spectro-temporal characteristics and compare them with properties of bursts of FRB~121102 detected at other observatories. We report on a precursor burst that is separated by just 17~ms from the main burst, the shortest reported separation between two individual bursts to date. We also provide access to data for all the detections of FRB~121102 in this campaign.

\end{abstract}

\section{Spectro-Temporal Analysis} \label{sec:intro}

\begin{figure}

\includegraphics[scale=0.64]{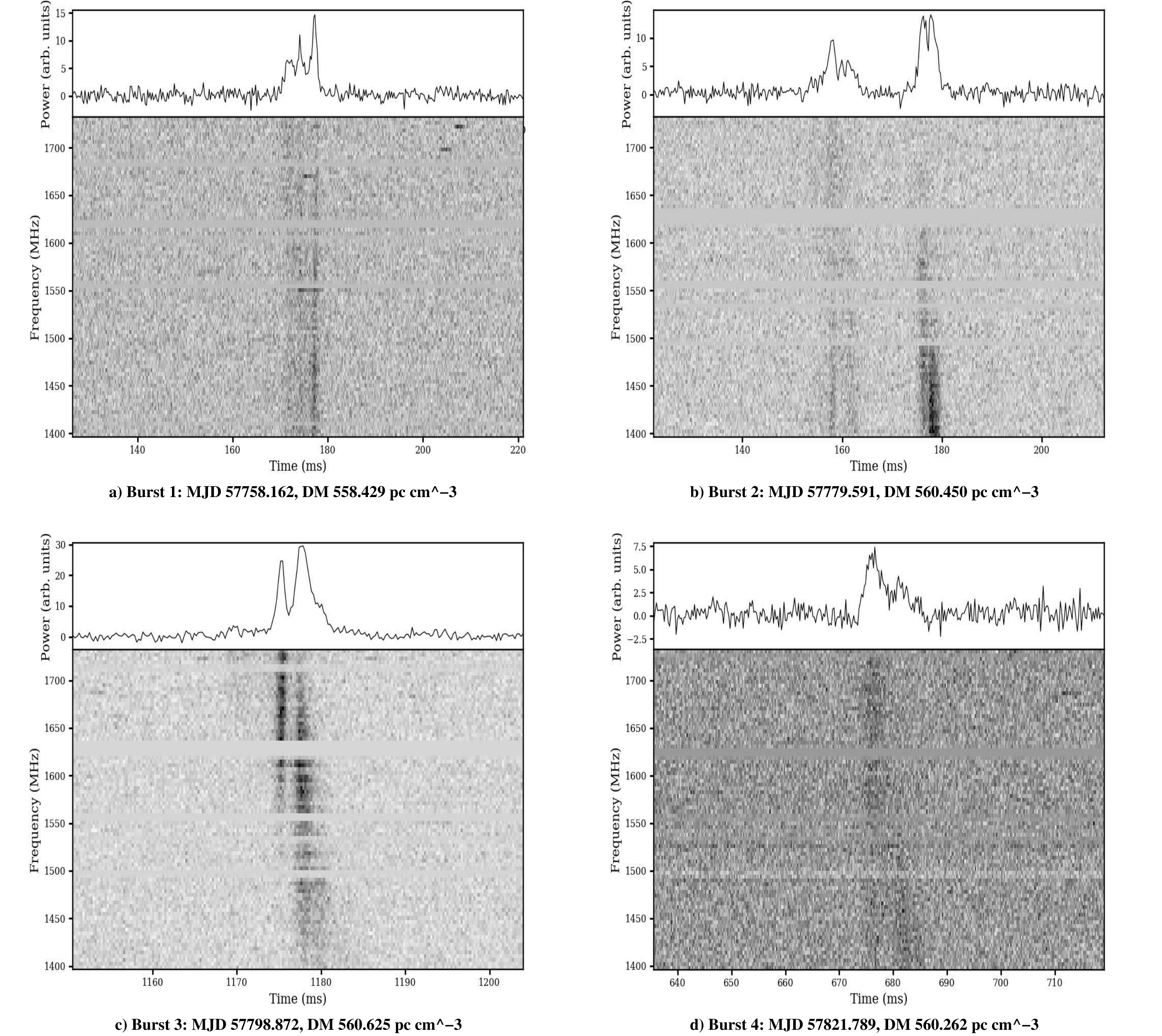}
\caption{Four bursts from FRB~121102 from the Lovell observing campaign. Each burst exhibits different emission characteristics in the dynamic spectrum. The bursts have been dedispersed to a DM that maximized the spectro-temporal structure.} 
\label{fig:r1_pulses}
\end{figure}

The results presented here are from data taken over the last five years as a part of an FRB~121102 monitoring program using the 76-m Lovell radio telescope at Jodrell Bank Observatory. The details of the observations and data processing are presented in~\cite{rajwade2020}. In this observing campaign, we detected 32 bursts from the source (more details of the bursts including widths and fluences are presented in~\cite{rajwade2020}). Some bursts from FRB~121102 are known to show unique emission characteristics in time and radio frequency. A number of authors have previously tried to quantify this behaviour~\citep[see][and the references therein]{Hessels}. So far, the most common trait observed is the presence of multiple sub-bursts within a single event that drift downwards and to later times as a function of observing frequency (the most obvious example in our sample can be seen in panel c in Figure \ref{fig:r1_pulses}). In order to investigate similar spectro-temporal characteristics, we selected bursts whose signal-to-noise ratio was high enough to reveal finer time-frequency structure at our minimum time resolution of 256~$\upmu$s. We visually inspected this sample and four bursts were found to show unique time-frequency characteristics. Firstly, we dedispersed the bursts to optimize the structure  as opposed to the signal-to-noise. In order to achieve this, we used the forward derivative of the burst as a parameter to quantify structure. The dispersion measure (DM) which maximsed this parameter was then used to dedisperse the data~\citep[see][for details]{gajjar2018}.  We discuss each of the four bursts (see Figure~\ref{fig:r1_pulses}) in the section below.

\section{burst Structure} \label{sec:style}

\subsection{Burst 1}
This burst shows three different components and unlike what has been seen for most drifting sub-bursts, all three components span the entire frequency band. The trailing component is observed to be brighter at the lower end of the band. The burst characteristics resemble those of single pulses seen in pulsars where multiple components of the same pulse span the entire band~\citep{beskin2015}. These type of bursts are not very commonly observed from FRB~121102 where multiple components (sub-bursts) are broadband with no evidence of drifting. Typically, broadband bursts exhibited by FRB~121102 are observed to have a single component. 

\subsection{Burst 2}
This is the most intriguing example in our sample. Two distinct bursts are seen here with the leading burst being fainter than the main burst as seen in other examples of precursor burst from FRB~121102~\citep{Hardy,caleb2020}. However in this case the leading burst is only fainter by a factor of$\sim$2 whereas other precursors seen from the source tend to be much fainter~\citep{Hardy, caleb2020}. The separation between the precursor and the main burst is $\sim$17~ms. This is the shortest of any reported separation between two distinct bursts in this source; so far they all tend to be separated by 20-40~ms~\citep{cruces2020}. There is also spectro-temporal structure evident in both the precursor and the main burst. There are three components in the precursor with the trailing component emerging in the lower part of the band similar to burst 1 while the main burst shows two components. It is unclear whether these are: two completely different bursts, a precursor-main burst scenario like seen previously, or bursts that are part of the same emission envelope. We note that burst B31 from \cite{cruces2020} is 39\,ms wide and those authors speculate it might represent the full pulse envelope, although B31 does exhibit a couple of components.

\subsection{Burst 3}
The sub-burst drifting seen in this example is like those more commonly seen in FRB~121102 bursts; with two sub-bursts clearly visible across the observing band. The sub-bursts span $\sim$300~MHz which is consistent with sub-bursts seen at different telescopes at 1.4~GHz~\citep{Hessels}. The leading component of the burst appears to broaden at lower frequencies as it gets fainter but the lack of finer time resolution prevents confirmation of the same. 

\subsection{Burst 4}
This burst clearly shows the emergence of a second component in the lower half of the band. The sub-burst behaviour looks different compared to the drifting sub-bursts seen from the Arecibo telescope at 1.4~GHz~\citep{Hessels} and resembles the trailing component of the precursor seen in burst 2. The burst also shares some similarities with burst 11 in the sample of bursts from the wide-band MeerKAT telescope study~\cite{caleb2020}. We do note that the widest bandwidth sub-burst seen at 1.4~GHz spans 400~MHz~\citep{Hessels} and a wider observing bandwidth would have led to a better classification of this burst.

\section{Summary} \label{sec:floats}
We have presented a sample of bursts from FRB~121102 that exhibit a rich variety in their spectro-temporal characteristics. The diversity can be attributed to a number of factors including propagation effects like plasma-lensing~\citep{main2018} and the intrinsic emission mechanism of the source~\citep{wang2019}. The DMs of the bursts optimized for structure are consistent with those measured from bursts detected with the Arecibo Telescope around the same time. The diversity in these bursts shows the importance of continued follow-up of repeating FRBs to collect a statistically-significant sample in order to provide stringent constraints on their origins. We provide the raw data for all of the bursts that were detected in the Lovell Telescope observations. The data are available for download in \textsc{ZENODO} at \dataset[10.5281/zenodo.3974768]{https://doi.org/10.5281/zenodo.3974768}.

\acknowledgments
KR, BS and MC acknowledge funding from the European Research Council's Horizon 2020 programme (grant agreement No.694745). RB acknowledges support from the European Research Council under the European Union's Horizon 2020 research and innovation programme (grant agreement no. 715051; Spiders). The authors would like to thank Andrew Lyne for help with the Lovell Telescope observations during this campaign.

\bibliography{refs}{}

\begin{thebibliography}{}
\expandafter\ifx\csname natexlab\endcsname\relax\def\natexlab#1{#1}\fi
\providecommand{\url}[1]{\href{#1}{#1}}
\providecommand{\dodoi}[1]{doi:~\href{http://doi.org/#1}{\nolinkurl{#1}}}
\providecommand{\doeprint}[1]{\href{http://ascl.net/#1}{\nolinkurl{http://ascl.net/#1}}}
\providecommand{\doarXiv}[1]{\href{https://arxiv.org/abs/#1}{\nolinkurl{https://arxiv.org/abs/#1}}}

\bibitem[{{Beskin} {et~al.}(2015){Beskin}, {Chernov}, {Gwinn}, \&
  {Tchekhovskoy}}]{beskin2015}
{Beskin}, V.~S., {Chernov}, S.~V., {Gwinn}, C.~R., \& {Tchekhovskoy}, A.~A.
  2015, \ssr, 191, 207, \dodoi{10.1007/s11214-015-0173-8}

\bibitem[{{Caleb} {et~al.}(2020){Caleb}, {Stappers}, {Abbott}, {Barr},
  {Bezuidenhout}, {Buchner}, {Burgay}, {Chen}, {Cognard}, {Driessen}, {Fender},
  {Hilmarsson}, {Hoang}, {Horn}, {Jankowski}, {Kramer}, {Lorimer}, {Malenta},
  {Morello}, {Pilia}, {Platts}, {Possenti}, {Rajwade}, {Ridolfi}, {Rhodes},
  {Sanidas}, {Serylak}, {Spitler}, {Townsend}, {Weltman}, {Woudt}, \&
  {Wu}}]{caleb2020}
{Caleb}, M., {Stappers}, B.~W., {Abbott}, T.~D., {et~al.} 2020, \mnras, 496,
  4565, \dodoi{10.1093/mnras/staa1791}

\bibitem[{{Cruces} {et~al.}(2020){Cruces}, {Spitler}, {Scholz}, {Lynch},
  {Seymour}, {Hessels}, {Gouiff{\`e}s}, {Hilmarsson}, {Kramer}, \&
  {Munjal}}]{cruces2020}
{Cruces}, M., {Spitler}, L.~G., {Scholz}, P., {et~al.} 2020, arXiv e-prints,
  arXiv:2008.03461.
\newblock \doarXiv{2008.03461}

\bibitem[{{Gajjar} {et~al.}(2018){Gajjar}, {Siemion}, {Price}, {Law},
  {Michilli}, {Hessels}, {Chatterjee}, {Archibald}, {Bower}, {Brinkman},
  {Burke-Spolaor}, {Cordes}, {Croft}, {Enriquez}, {Foster}, {Gizani},
  {Hellbourg}, {Isaacson}, {Kaspi}, {Lazio}, {Lebofsky}, {Lynch}, {MacMahon},
  {McLaughlin}, {Ransom}, {Scholz}, {Seymour}, {Spitler}, {Tendulkar},
  {Werthimer}, \& {Zhang}}]{gajjar2018}
{Gajjar}, V., {Siemion}, A.~P.~V., {Price}, D.~C., {et~al.} 2018, \apj, 863, 2,
  \dodoi{10.3847/1538-4357/aad005}

\bibitem[{{Hardy} {et~al.}(2017){Hardy}, {Dhillon}, {Spitler}, {Littlefair},
  {Ashley}, {De Cia}, {Green}, {Jaroenjittichai}, {Keane}, {Kerry}, {Kramer},
  {Malesani}, {Marsh}, {Parsons}, {Possenti}, {Rattanasoon}, \&
  {Sahman}}]{Hardy}
{Hardy}, L.~K., {Dhillon}, V.~S., {Spitler}, L.~G., {et~al.} 2017, \mnras, 472,
  2800, \dodoi{10.1093/mnras/stx2153}

\bibitem[{{Hessels} {et~al.}(2019){Hessels}, {Spitler}, {Seymour}, {Cordes},
  {Michilli}, {Lynch}, {Gourdji}, {Archibald}, {Bassa}, {Bower}, {Chatterjee},
  {Connor}, {Crawford}, {Deneva}, {Gajjar}, {Kaspi}, {Keimpema}, {Law},
  {Marcote}, {McLaughlin}, {Paragi}, {Petroff}, {Ransom}, {Scholz}, {Stappers},
  \& {Tendulkar}}]{Hessels}
{Hessels}, J.~W.~T., {Spitler}, L.~G., {Seymour}, A.~D., {et~al.} 2019, \apjl,
  876, L23, \dodoi{10.3847/2041-8213/ab13ae}

\bibitem[{{Main} {et~al.}(2018){Main}, {Yang}, {Chan}, {Li}, {Lin}, {Mahajan},
  {Pen}, {Vanderlinde}, \& {van Kerkwijk}}]{main2018}
{Main}, R., {Yang}, I.~S., {Chan}, V., {et~al.} 2018, \nat, 557, 522,
  \dodoi{10.1038/s41586-018-0133-z}

\bibitem[{{Rajwade} {et~al.}(2020){Rajwade}, {Mickaliger}, {Stappers},
  {Morello}, {Agarwal}, {Bassa}, {Breton}, {Caleb}, {Karastergiou}, {Keane}, \&
  {Lorimer}}]{rajwade2020}
{Rajwade}, K.~M., {Mickaliger}, M.~B., {Stappers}, B.~W., {et~al.} 2020,
  \mnras, 495, 3551, \dodoi{10.1093/mnras/staa1237}

\bibitem[{{Wang} {et~al.}(2019){Wang}, {Zhang}, {Chen}, \& {Xu}}]{wang2019}
{Wang}, W., {Zhang}, B., {Chen}, X., \& {Xu}, R. 2019, \apjl, 876, L15,
  \dodoi{10.3847/2041-8213/ab1aab}

\end{thebibliography}
\bibliographystyle{aasjournal}

\end{document}